\documentclass[aps,reprint,prl,showpacs,superscriptaddress]{revtex4-2}
\usepackage{amsmath}
\usepackage{amssymb}
\usepackage{graphicx}
\usepackage{color}
\usepackage{bm}
\usepackage{siunitx}
\usepackage{mathtools}
\usepackage{MnSymbol}
\usepackage{cancel}

\usepackage{babel}

\newcommand{\nn}{\langle\bm{n}\bm{n}\rangle}
\newcommand{\nnnn}{\langle\bm{nnnn}\rangle}
\newcommand{\phiRCP}{\phi_\mathrm{RCP}}
\newcommand{\phiJ}{\phi_\mathrm{J}}

\DeclareMathOperator{\Tr}{Tr}
\DeclareMathOperator{\sgn}{sgn}

\graphicspath{ {./figs/} }

\begin{document}

\title{Rheology of dense suspensions under shear rotation}

\author{Frédéric Blanc}
\author{François Peters}
\affiliation{Université Côte d'Azur, CNRS, Institut de Physique de Nice (INPHYNI), France}

\author{Jurriaan J.J. Gillissen}
\affiliation{The Technology Partnership,
Science Park, Melbourn, UK.}

\author{Michael E. Cates}
\affiliation{DAMTP, Centre for Mathematical Sciences, University of Cambridge, Wilberforce Road, Cambridge CB3 0WA, UK}

\author{Sandra Bosio}
\author{Camille Benarroche}
\affiliation{Université Côte d'Azur, CNRS, Institut de Physique de Nice (INPHYNI), France}

\author{Romain Mari}
\affiliation{Univ. Grenoble Alpes, CNRS, LIPhy, 38000 Grenoble, France}

\date{\today}

\begin{abstract}
Dense non-Brownian suspensions exhibit a spectacular and abrupt drop in viscosity under change of shear direction, as revealed by shear inversions (reversals) or orthogonal superposition. 
Here, we introduce an experimental setup to systematically explore their  response to shear rotations, where one suddenly rotates the principal axes of shear by an angle $\theta$, and measure the shear stresses with a bi-axial force sensor. 
Our measurements confirm the genericness of the transient decrease of the resistance to shear under unsteady conditions.
Moreover, the orthogonal shear stress, which vanishes in steady state, takes non-negligible values with a rich $\theta$-dependence, 
changing qualitatively with solid volume fraction $\phi$, and resulting in a force that tends to reduce or enhance the direction of flow for small or large $\phi$.   
These experimental findings are confirmed and rationalized by particle-based numerical simulations and a recently proposed constitutive model.
We show that the rotation angle dependence of the orthogonal stress results from a $\phi$-dependent interplay between hydrodynamic and contact stresses.
\end{abstract}

\maketitle

Suspensions of non-Brownian hard particles 
form a large class of complex fluids~\cite{denn_rheology_2014,ness_physics_2022}.
They are dense when solid and fluid are mixed in roughly equal proportion.
Their widespread use in industry calls for  proper constitutive characterization and modelling to enable reliable process design.

In steady state, their viscosity is either deformation-rate-independent~\cite{ovarlez_local_2006,boyer_unifying_2011,guyUnifiedDescriptionRheology2015} or slightly shear thinning~\cite{zarraga_characterization_2000,dai_viscometric_2013,dbouk_normal_2013,vazquez-quesadaShearThinningNoncolloidal2016,lobry_shear_2019}. 
Shear thickening is also observed when particles are repulsive beyond pure hard-core forces~\cite{setoDiscontinuousShearThickening2013,wyartDiscontinuousShearThickening2014}, but we here focus on the strictly hard-sphere case.
The viscosity increases with the solid volume fraction $\phi$, and diverges at the jamming volume fraction $\phi_\mathrm{J}$~\cite{ovarlez_local_2006,boyer_unifying_2011}.
However, dense suspensions exhibit striking unsteady behaviors,
e.g. the sharp viscosity drop in orthogonal superposition~\cite{ovarlez2010three,barral2011superposition,blanc_tunable_2014,lin_tunable_2016,ness_shaken_2018,ramaswamyIncorporatingTunabilityUniversal2022,acharyaOptimumDissipationCruising2023}.
In shear reversal, where a suspension initially sheared in steady state under a deformation rate $\dot \gamma$ is suddenly sheared with a rate $-\dot\gamma$, the viscosity drops suddenly at reversal, passes through a minimum value and climbs up to its steady-state value after a few strain units~\cite{gadalamaria_shearinduced_1980,narumi_transient_2002,kolli_transient_2002,blanc_local_2011,linHydrodynamicContactContributions2015,peters_rheology_2016,nessTwoscaleEvolutionShear2016,chackoShearReversalDense2018}.

Under shear, suspensions develop an anisotropic micro-structure which takes up most of the stress at large concentrations~\cite{lootensDilatantFlowConcentrated2005,blancExperimentalSignaturePair2011,blancMicrostructureShearedNonBrownian2013,setoDiscontinuousShearThickening2013,mariShearThickeningFrictionless2014,gallierRheologyShearedSuspensions2014,guyUnifiedDescriptionRheology2015,linHydrodynamicContactContributions2015} and is built in a finite strain~\cite{gadalamaria_shearinduced_1980}.
Upon shear reversal, the micro-structure is initially not compliant with the new direction of shear, 
leading to a viscosity dip which ends when 
the micro-structure is rebuilt in the new orientation~\cite{gadalamaria_shearinduced_1980}.
This behavior may be a vestige of the fragility of jammed suspensions that can be made to flow by a change of applied load direction~\cite{cates_jamming_1998,setoShearJammingFragility2019,giusteriShearJammingFragility2021}.

Characterization of the mechanical response to sudden changes of the strain axes, beyond shear reversal (which is the extreme case, as the compressional and elongational axes are swapped)~\cite{peters_rheology_2016,nessTwoscaleEvolutionShear2016}, is however absent.
Here we fill this gap by considering the response to \emph{shear rotations}, i.e. rotations of the strain axes by an arbitrary angle $\theta$ about the gradient direction. 
We perform shear rotations in experiments, with a specifically designed rheometer; simulations, using discrete element method (DEM); and the Gillissen-Wilson (GW) constitutive model~\cite{gillissen_modeling_2018,gillissen_constitutive_2020}.
This gives us access to the viscosity drop 
as a function of angle $\theta$ and post-rotation strain $\gamma$.
This also unveils a new non-Newtonian phenomenon: following a shear rotation, the shear viscosity orthogonal to the flow direction, $\eta_{32}$, is transiently finite, reaching up to \SI{50}{\percent} of the usual shear viscosity $\eta_{12}$ for large $\phi$.
Moreover, we show that the nature of angular dependence of $\eta_{32}$ depends on $\phi$: while at moderate $\phi$, $\eta_{32}$ shows a change of sign in $\theta \in [0, \pi]$, associated to a force resisting shear rotation for small $\theta$ values, for the largest $\phi$ values it keeps a constant sign.
We show that this is due to the decreasing relative contribution of hydrodynamic stresses versus contact stresses when $\phi$ increases.


\paragraph{Experimental setup}

\begin{figure}[t]
\centering
\includegraphics[width=\columnwidth]{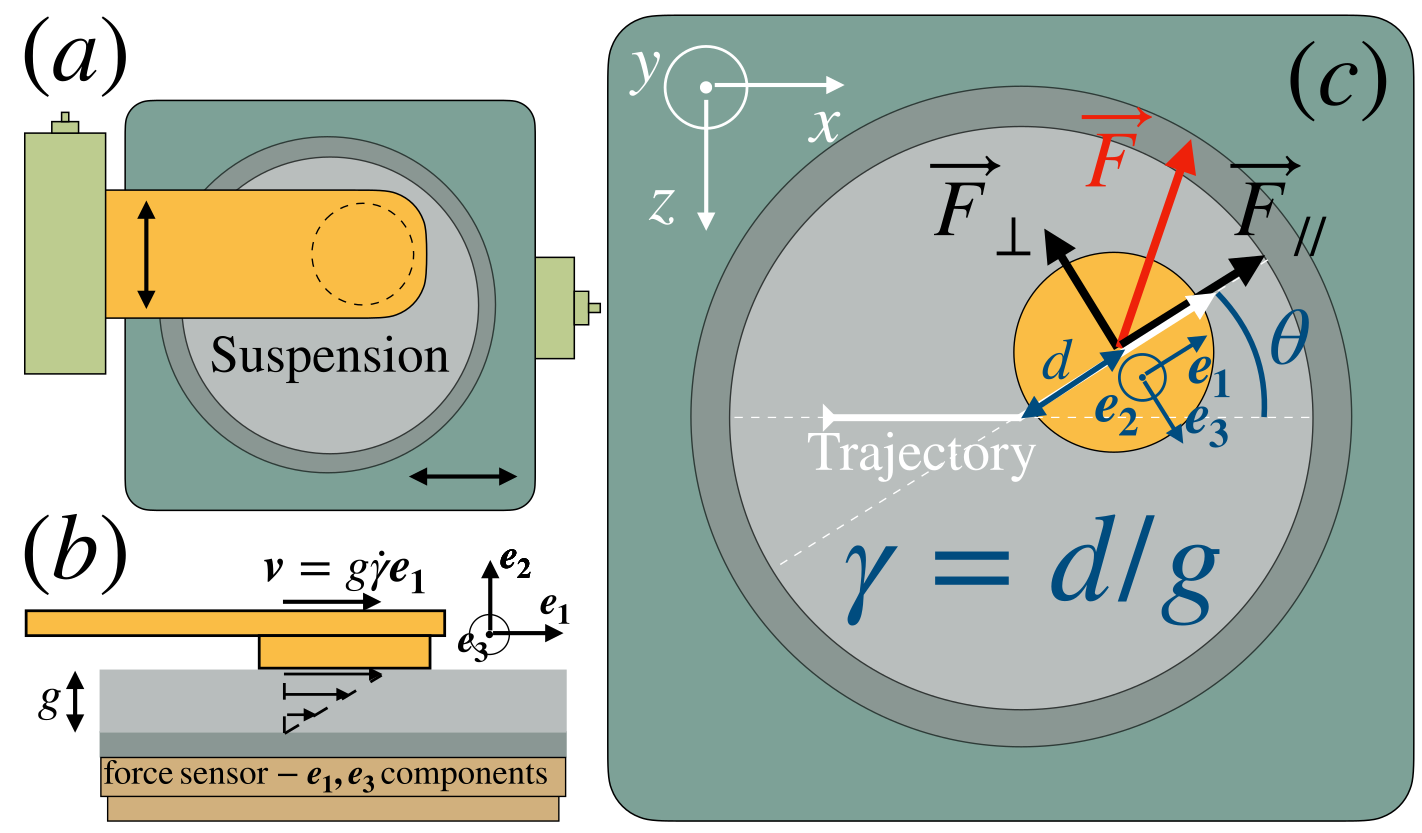}
\caption{Sketch of the setup. (a) View along the flow-gradient direction. Two translation stages (light green) independently move an upper plate (orange) and a lower plate (green gray), between which the suspension (light 
gray) is sheared. (b) Shear plane view. A force sensor 
measures the tangential stresses. 
(c) View along the flow-gradient direction of the trajectory (thick white line) of the top plate relative to the bottom plate during a shear rotation by an angle $\theta$. A transient force $\bm{F} = F_{\parallel}\bm{e}_1 + F_{\perp}\bm{e}_3$ is recorded after the shear rotation.}
\label{fig:setup}
\end{figure}

We designed a \emph{cross rheometer}~\cite{linMultiaxisConfocalRheoscope2014,SI}, sketched in Fig.~\ref{fig:setup}, made with two parallel plates mounted on two motorized linear stages of 25mm stroke (Newport MFA-CC) acting in perpendicular directions, allowing for arbitrary relative parallel motion and therefore arbitrary simple shear with velocity gradient orthogonal to the plates. 
We apply a simple shear (Fig.~\ref{fig:setup}(b)) with velocity gradient $\bm{L} = \dot\gamma \bm{e}_1 \bm{e}_2$ (with $\bm{e}_1$, $\bm{e}_2$ and $\bm{e}_3$ respectively the flow, gradient and vorticity directions), from which we define the strain-rate tensor $\bm{E} =\dot\gamma \hat{\bm{E}} \equiv (\bm{L} + \bm{L}^\mathrm{T})/2$. 
The shear rate $\dot\gamma$ is related to the velocity of the top plane relative to the bottom plane $\bm{v} = \dot\gamma g \bm{e}_1$, $g=\SI{1}{\milli\meter}$ being the gap width between the two plates. 

A force sensor (AMTI HE6x6-1) measures the tangential force $\vec{F}=\vec{F}_\parallel + \vec{F}_\perp$ exerted on the lower plate (and thus by the upper plate on the suspension), with $\vec{F}_\parallel$ and $\vec{F}_\perp$ respectively along $\bm{e}_1$ and $\bm{e}_3$ (Fig.~\ref{fig:setup}(b)-(c)).
We define the shear viscosity $\eta_{12} \equiv \bm{\Sigma}:\bm{e}_\mathrm{1}\bm{e}_\mathrm{2}/\dot\gamma = \vec{F}_\parallel \cdot \bm{e}_1/(S\dot\gamma)$, with $S$ the area of the upper plate, as well as the ``orthogonal'' shear viscosity $\eta_{32} \equiv \bm{\Sigma}:\bm{e}_\mathrm{3}\bm{e}_\mathrm{2}/\dot\gamma = \vec{F}_\perp \cdot \bm{e}_3/(S\dot\gamma)$. 
Both have opposite $\theta$-parities: $\eta_{12}$ is even, $\eta_{12}(\gamma, \theta) = \eta_{12}(\gamma, -\theta)$, while $\eta_{32}$ is odd, $\eta_{32}(\gamma, \theta) = -\eta_{32}(\gamma, -\theta)$ (in particular this enforces that $\eta_{32}(\gamma, 0)$ and $\eta_{32}(\gamma, \pi)$ vanish). 

A preshear is first applied over 10 strain units, which is enough to reach steady state, followed  by a shear rotation where we rotate the flow direction $\bm{e}_1$ and vorticity direction $\bm{e}_3$  by an angle $\theta \in [-\pi, \pi]$ around the gradient direction $\bm{e}_2$ (thick white line in Fig. \ref{fig:setup}(c)). The imposed shear rate is set at $\dot{\gamma} = 0.4 \, \mathrm{s}^{-1}$ for all presented experimental data.

The viscosities $\eta_{12}(\gamma,\theta)$ and $\eta_{32}(\gamma,\theta)$ are recorded via a Data Acquisition System (USB-1608FS-PLUS, MCCDAQ) as a function of the subsequent strain $\gamma < 10$.
The strain resolution is $\approx \num{3e-3}$, the lowest available strain is $\approx \num{1e-2}$, and $\theta$ is sampled every $\pi/18$.

The suspension particles are polystyrene spheres with a diameter of $\SI{40}{\micro\meter}$ (Microbeads TS40). They are dispersed in poly(ethylene glycol-ran-propylene glycol) (Sigma-Aldrich,  viscosity $\eta_0=\SI{38.4}{\pascal\second}$ at $\SI{25}{\celsius}$, $M_\mathrm{n} \approx \num{12000}$) for suspensions at $\phi = 0.45$ or silicone oil (M1000, Roth, $\eta_0 = \SI{0.98}{\pascal\second}$ at $\SI{25}{\celsius}$)  at $\phi=0.55$ and $0.57$.

\paragraph{Numerics \& model}

We perform the same protocol in DEM simulations of a suspension of $N=2000$ frictional particles subject to lubrication and contact forces in a tri-periodic configuration, using a method described in~\cite{SI,mariShearThickeningFrictionless2014}. 
The suspension is bidisperse (with a size ratio $1:1.4$) and the friction coefficient is $\mu_\mathrm{p} = 0.5$, in order to  match the viscosity values observed experimentally.

We also compare our results to the predictions of the GW model, a model capturing the features of shear reversal~\cite{gillissen_modeling_2018,gillissenConstitutiveModelTimeDependent2019}.
A detailed derivation and discussion of the underlying assumptions of the GW model can be found in \cite{gillissen_constitutive_2020}.
The GW model considers the strain evolution of a fabric tensor $\nn$ where $\bm{n}$ is the unit separation vector between pairs of particles in a near interaction via contact or lubrication forces
\begin{multline}
\partial_\gamma\nn =
\hat{\bm{L}}\cdot\nn+\nn \cdot\hat{\bm{L}}^\mathrm{T}-
2\hat{\bm{L}}:\nnnn \\
-\beta\left[
\hat{\bm{E}}_\mathrm{e}: \nnnn+
\frac{
\color{black}
\phi
\color{black}
}{15}
 \left(2\hat{\bm{E}}_\mathrm{c}+\Tr(\hat{\bm{E}}_\mathrm{c})
 \bm{\delta}\right)
 \right]\, .
  \label{eq:GW_fabric}
\end{multline}
The top line of Eq. (\ref{eq:GW_fabric}) describes that particle pairs rotate like dumbbells with the velocity gradient $\hat{\bm{L}}=\bm{L}/\dot\gamma$ and the bottom line describes the association and dissociation of particle pairs due to the compressive part $\hat{\bm{E}}_\mathrm{c}$ and the extensive part $\hat{\bm{E}_\mathrm{e}}$ of the strain rate tensor $\hat{\bm{E}}$ which pushes particles together and pulls them apart, respectively. 
The pair association (and dissociation) rate $\beta$ is a tuneable parameter.
The tensor $\nnnn$ is approximated in terms of $\nn$ with the Hinch \& Leal closure~\cite{hinch1976constitutive}.
Furthermore, the GW model decomposes the stress $\bm{\Sigma} = \bm{\Sigma}^\mathrm{H}+\bm{\Sigma}^\mathrm{C} + 2\eta_0 \bm{E}$ in contributions from hydrodynamics, $\bm{\Sigma}^\mathrm{H}$, and contacts, $\bm{\Sigma}^\mathrm{C}$, 
\begin{align}
\frac{\bm{\Sigma}^\mathrm{H}}{\eta_\mathrm{s}\dot\gamma} = 
\frac{\alpha_0 \hat{\bm{E}} :\nnnn}{\left(1-\phi/\phiRCP\right)^{2}}, 
&&
\frac{\bm{\Sigma}^\mathrm{C}}{\eta_\mathrm{s}\dot\gamma} = 
\frac{\chi_0 \hat{\bm{E}}_\mathrm{c}
 :\nnnn}{\left(1-\xi/\xi_\mathrm{J}\right)^{2}},  
\label{eq:GW_stress}
\end{align}
where $\alpha_0$ and $\chi_0$ are tuneable parameters. $\bm{\Sigma}^\mathrm{H}$ diverges when $\phi$ approaches the random close packing volume fraction $\phiRCP=0.65$ and $\bm{\Sigma}^\mathrm{C}$ diverges when the `jamming coordinate' $\xi=-\nn:\bm{E}_\mathrm{c} \vert\bm{E}_\mathrm{c}\vert^{-1}$, a proxy for the coordination number,  approaches 
the  jamming value $\xi_\mathrm{J}$. By demanding that, 
in steady shear flow, $\bm{\Sigma}^\mathrm{C}$ diverges when 
$\phi$ approaches the friction-dependent jamming volume fraction $\phiJ=0.58$, we have previously shown that:
$\xi_\mathrm{J}=\phiJ\left(213 \beta ^2-234 \beta +2080\right) \left[ 15 \left(9 \beta ^2+54 \beta +416\right)\right]^{-1}$~\cite{gillissen_constitutive_2020}. 
In the SI we argue our choices for $\alpha_0=2.4$, $\chi_0=2.3$ and $\beta=7$.

\paragraph{Results}

\begin{figure}[t]
\centering
\includegraphics[width=\columnwidth]{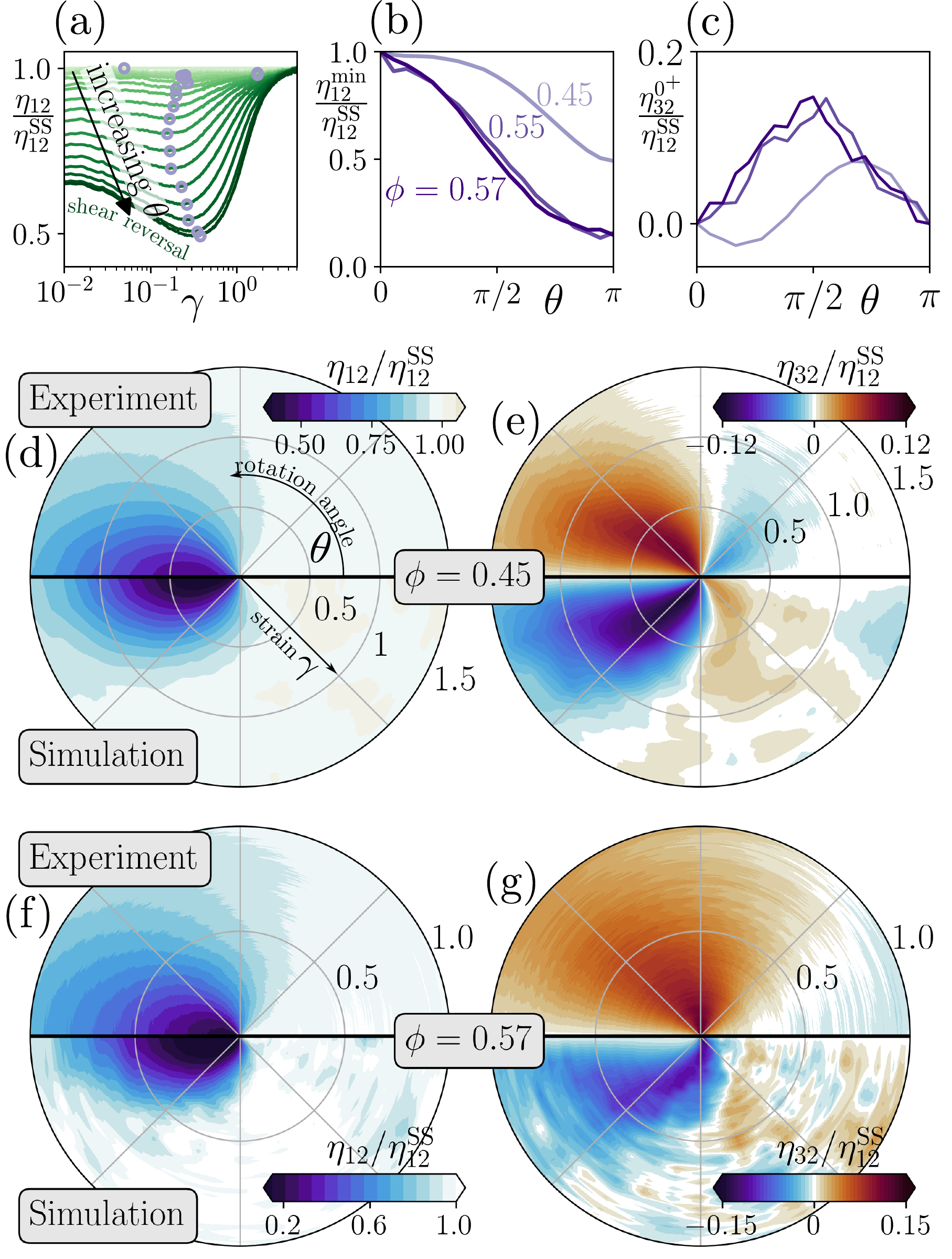}
\caption{(a) Shear viscosity $\eta_{12}$ normalized by the steady-state value $\eta_{12}^\mathrm{SS}$ as a function of strain $\gamma$ in experiments for $\phi=0.45$, for several values of $\theta \in [0, \pi]$, increasing from light to dark.
The minimum values of the viscosity $\eta_{12}^{\mathrm{min}}/\eta_{12}^\mathrm{SS}$ for each $\theta$ are circled.
(b) and (c) $\eta_{12}^{\mathrm{min}}/\eta_{12}^\mathrm{SS}$ and orthogonal viscosity just after rotation $\eta_{32}^{0^+}/\eta_{12}^\mathrm{SS}$ as a function of shear rotation angle $\theta$ for $\phi=0.45, 0.55$ and $0.57$.
(d) and (e) Polar representation of $\eta_{12}(\gamma, \theta)$ and $\eta_{32}(\gamma, \theta)$, normalized by $\eta_{12}^\mathrm{SS}$, for $\phi=0.45$, in experiments (top halves) and numerical simulations (bottom halves). Symmetries impose that $\eta_{12}$ is even, $\eta_{12}(\gamma, \theta) = \eta_{12}(\gamma, -\theta)$, and $\eta_{32}$ is odd, $\eta_{32}(\gamma, \theta) = -\eta_{32}(\gamma, -\theta)$.
(f) and (g) Same, but for $\phi=0.57$.}
\label{fig:radial_comparison}
\end{figure}

In Fig.~\ref{fig:radial_comparison}(a), we show the viscosity $\eta_{12}(\gamma, \theta)$ measured in experiments, 
for a moderately dense suspension at $\phi = 0.45$. 
It decreases at low strain values, then passes through a minimum before increasing back to its steady-state value. 
The minimum is located at a strain $\gamma_\mathrm{min}$ weakly dependent on $\theta$, from $\gamma_\mathrm{min} \approx 0.15$ for $\theta\approx \pi/2$ to $\gamma_\mathrm{min} \approx 0.35$ for shear reversal ($\theta=\pi$).
As shown in Fig.~\ref{fig:radial_comparison}(b), the minimum value $\eta_{12}^{\mathrm{min}}$ gradually decreases when $\theta$ increases, to reach its lowest value for shear reversal.
Once normalized by the steady-state value $\eta_{12}^\mathrm{SS}$, $\eta_{12}^{\mathrm{min}}/\eta_{12}^\mathrm{SS}$ for a given $\theta$ decreases when $\phi$ increases, as is already known for shear reversal~\cite{peters_rheology_2016}.
Interestingly, for the lowest $\phi=0.45$, $\eta_{12}^{\mathrm{min}} \approx \eta_{12}^\mathrm{SS}$ for $\theta \lesssim \pi/4$: 
the suspension seems oblivious to the shear rotation at small angles.
We will see that this is not quite true when considering $\eta_{32}$.

We compare these data with the DEM ones in a radial representation $\eta_{12}(\gamma, \theta)$ in Fig.~\ref{fig:radial_comparison}(d), with experiments in the top half and numerics in the bottom half. 
The agreement is good, besides simulations predicting a quicker relaxation to steady state than actually observed.

We turn in Fig.~\ref{fig:radial_comparison}(e) to $\eta_{32}$, again comparing experiments in the top half and numerics in the bottom half.
Both datasets are in excellent agreement and reveal a structure mixing first and second order odd circular harmonics (respectively $\propto \sin\theta$ and $\propto \sin 2\theta$) with similar amplitudes. 
For $0< \theta \lesssim \pi/2$, we find $\eta_{32} < 0$ for $\gamma \lesssim 1$, i.e. the suspension exerts on the top plate a ``restoring'' force in the direction of decreasing $\theta$ values.
In a force control setup where one sets the upper plate force $\vec{F}_\parallel$ rather than its displacement, the suspension would thus be stable with respect to shear rotations, by rotating the velocity of the top plate towards lower $\theta$ values. 
By contrast, for $\pi/2 \lesssim \theta < \pi$, we find $\eta_{32} > 0$ for $\gamma \lesssim 2$, which can be interpreted as the suspension tending to rotate the trajectory of the top plate towards larger $\theta$ values.

In Fig.~\ref{fig:radial_comparison}(f),(g), we show $\eta_{12}$ and $\eta_{32}$ for $\phi = 0.57$.
Both experimental and numerical data show that the relaxation to steady state is quicker than at $\phi=0.45$, with a smaller $\eta_{12}^{\mathrm{min}}/\eta_{12}^\mathrm{SS}$ value~\cite{peters_rheology_2016,chackoShearReversalDense2018}.
More importantly, the first harmonic of $\eta_{32}$ dominates. 
For $0<\theta<\pi$, we find $\eta_{32} > 0$: the suspension always tends to push the top plate to move towards larger $\theta$ values, that is, the response is no longer stabilizing for small shear rotations. In Fig.~\ref{fig:radial_comparison}(c), we highlight this qualitative change by showing the $\theta$ dependence of the values just after rotation $\eta_{32}^{0^+}$.

Our simulations also show that the fabric evolution after shear rotation  does not mimic the full stress response but only its contact contribution~\cite{SI}. 
Notably, the fabric evolution does not exhibit any qualitative change upon increase of $\phi$ that we could correlate to the change of behavior of $\eta_{32}$.

\begin{figure}[t]
\centering
\includegraphics[width=\columnwidth]{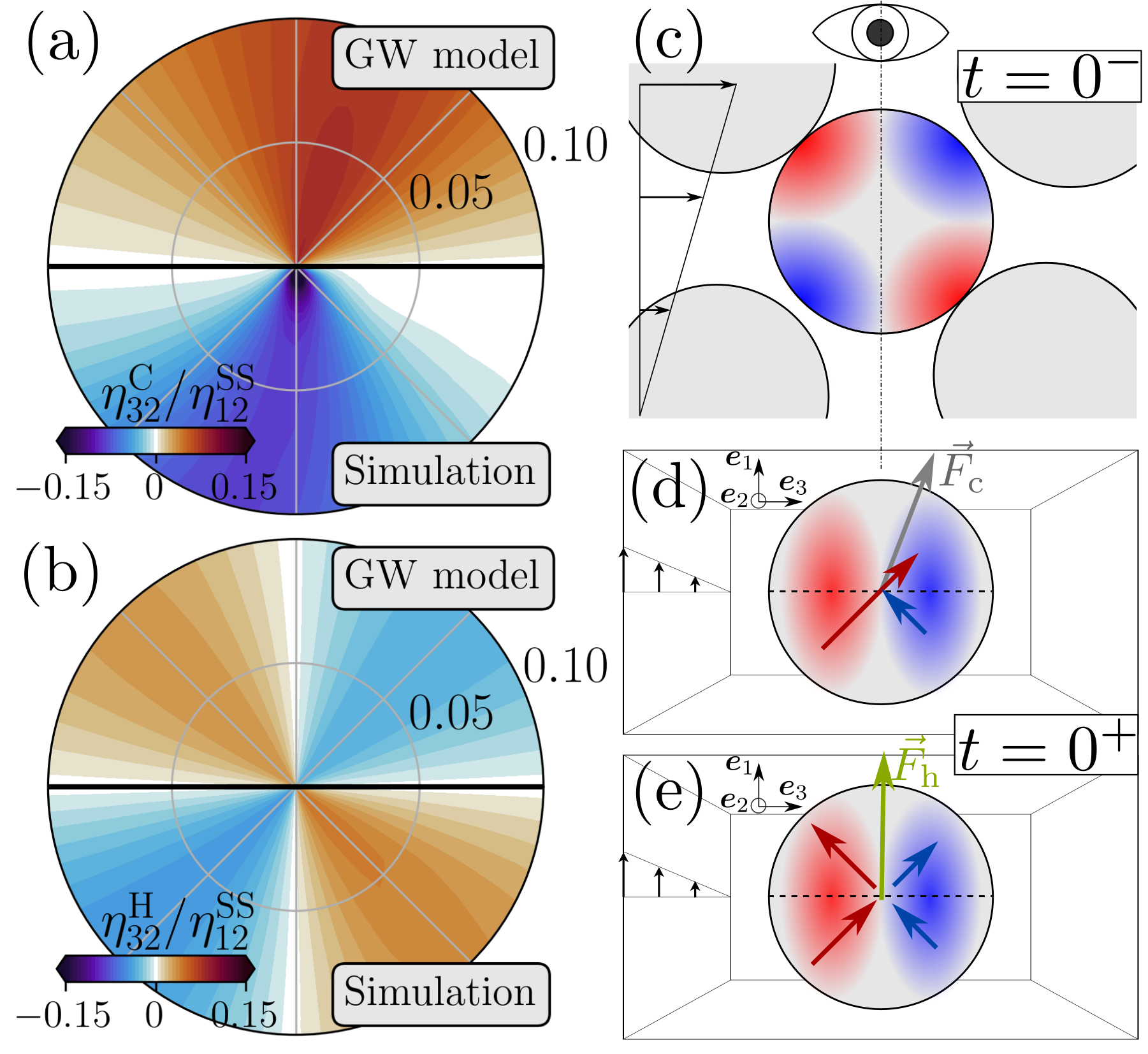}
\caption{Contributions to the orthogonal viscosity from contacts (a) and hydrodynamics (b), in the GW model in the top halves and DEM simulations in the bottom halves, for $\phi=0.45$. 
We chose $\phiRCP= 0.65$, $\phiJ = 0.58$,
$\alpha_0=2.3$
$\chi_0=2.4$, and $\beta=7$, based on earlier comparisons with DEM simulations~\cite{gillissen_constitutive_2020,SI}. 
We recall that $\eta_{32}$ is odd, $\eta_{32}(\gamma, \theta) = -\eta_{32}(\gamma, -\theta)$. (c) A particle during initial shear at $t=0^-$ has more near contacts in compressional quadrants (red) than elongational quadrants (blue). (d)--(e) Looking down from the gradient direction, just after shear rotation by $\theta=\pi/2$ the new compressional and elongational quadrants are respectively below and above the dashed lines. Contact forces come from the new compressional, and are dominated by the more numerous contacts in the old compressional (red arrow), leading to $\eta_{32}^\mathrm{C}>0$ (d). Hydrodynamic forces have symmetric contributions from new compressional and elongational quadrants, leading to $\eta_{32}^\mathrm{H}=0$ (e).}
\label{fig:visc_components}
\end{figure}

To understand the origin of the $\eta_{32}$ behavior, we interrogate the GW model and its predictions for the contact and hydrodynamic contributions to the stress response.
As shown in Fig.~\ref{fig:visc_components}(a), the contact contribution $\eta^\mathrm{C}_{32}$ has a dominant first harmonic, with $\eta^\mathrm{C}_{32}>0$ for $0 < \theta < \pi$.
The large second harmonic of $\eta_{32}$ is instead due to the hydrodynamic component $\eta^\mathrm{H}_{32}$ (Fig.~\ref{fig:visc_components}(b)), 
which at $\phi=0.45$ still accounts for a substantial part of the total stress~\cite{gallierRheologyShearedSuspensions2014}. 
This is confirmed by numerical simulations, which compare well to the predictions of the model.

The difference between contact and hydrodynamic contributions can be qualitatively understood. 
In Fig.~\ref{fig:visc_components}(c), we sketch in the shear plane a particle during preshear.
It shows a fore-aft asymmetry: it has more near interactions (lubricated and in contact) in the compressional quadrants (in red) than in the elongational one (in blue).
The same particle is seen from the gradient direction $\bm{e}_2$ right after a shear rotation with $\theta = \pi/2$ in Fig.~\ref{fig:visc_components}(d),(e).
After rotation, the fore-aft asymmetry accumulated in preshear is a ``left-right'' asymmetry, and fore-aft symmetry is temporarily restored.
Post-rotation contact stresses (Fig.~\ref{fig:visc_components}(d)) stem from contacts in the post-rotation compressional quadrant, below the dashed line, 
and due to the left-right asymmetry, are dominated by contacts that carry over from the pre-rotation in the intersect of pre- and post-rotation compressional quadrants. 
Contact forces $\vec{F}_\mathrm{C}$ in this overlap region (red vector) are such that $\bm{e}_3\cdot \vec{F}_\mathrm{C}>0$, giving a positive contribution to $\eta_{32}$. 
By contrast, in Fig.~\ref{fig:visc_components}(d), all interactions contribute hydrodynamic forces, and the fore-aft symmetry ensures that the hydrodynamic contribution from the post-rotation elongational and compressional quadrants share the same $\bm{e}_1$ component, but have opposite $\bm{e}_3$ component. This results in a net-zero hydrodynamic contribution to $\eta_{32}$.

Whereas this reasoning can be extended to show that $\eta^\mathrm{C}_{32} > 0$ for $\theta \in ]0, \pi[$, the sign of $\eta^\mathrm{H}_{32}$ for $\theta \neq \pi/2$ depends on aspects of the distribution of near interactions that cannot be deduced from symmetry considerations.
The GW model however gives us a quantitative picture alongside a microstructural insight. 
Calling $\nn^\mathrm{ss}$ the steady-state fabric in pre-shear, we get the following contributions to $\eta_{32}$ at $\gamma=0^+$~\cite{SI}
\begin{align}
    \frac{\eta^\mathrm{H}_{32}}{\eta_\mathrm{s}} & = \frac{\alpha(\phi)}{14} \sin 2\theta \left(\nn^\mathrm{ss}_{xx} - \nn^\mathrm{ss}_{zz}\right) \label{eq:GW_contact_vs_hydro}
\\
    \frac{\eta^\mathrm{C}_{32}}{\eta_\mathrm{s}} & 
= \frac{\chi(\xi)}{7} \left[ \frac{\sin 2\theta}{4}  \left(\nn^\mathrm{ss}_{xx} - \nn^\mathrm{ss}_{zz}\right)- \sin\theta \nn^\mathrm{ss}_{xy}\right] \, ,
\nonumber 
\end{align}
with
\begin{multline*}
    \xi = \big[\cos^2 \theta \nn^\mathrm{ss}_{xx} + \nn^\mathrm{ss}_{yy} + \sin^2\theta \nn^\mathrm{ss}_{zz} \\ 
    - 2\cos\theta \nn^\mathrm{ss}_{xy}\big]/2\, .
\end{multline*}
For the contact contribution, the first harmonic dominates for the values of $\phi$ and $\beta$ investigated. 
It is such that $\eta^\mathrm{C}_{32}>0$ for $0 < \theta < \pi$.
By contrast, the hydrodynamic contribution only has a second harmonic. 
Therefore, when $\alpha(\phi)/\chi(\xi)$ is large enough, the four-lobed hydrodynamic response dominates, which happens at moderate $\phi$. 
 However, since contact stresses diverge for $\phi \to \phiJ$ 
 while hydrodynamic stresses diverge for $\phi \to \phiRCP > \phiJ$,
 the two-lobed contact response takes over when $\phi$ moves closer to $\phi_J$.

The sign of $\eta^\mathrm{H}_{32}$ is not obvious, as it is set by $\nn^\mathrm{ss}_{xx} - \nn^\mathrm{ss}_{zz}$.
With our value of $\beta=7$, we always find $\nn^\mathrm{ss}_{xx} - \nn^\mathrm{ss}_
{zz}<0$ in the GW model.
For small $\theta$ values, it therefore ``stabilizes'' the microstructure, as $\sgn \eta^\mathrm{H}_{32} = -\sgn \theta$: 
 it provides a restoring force acting against the rotation of the flow direction. 
In simulations, $\nn^\mathrm{ss}_{xx} - \nn^\mathrm{ss}_{zz}$ is measured tiny~\cite{chackoShearReversalDense2018}.
To test the GW model predictions, from our DEM simulations we compute $\nn$ based on particle pairs separated by at most a gap of $\epsilon_\mathrm{c} = 0.05$ times the average radius of the pair. 
We find $\nn^\mathrm{ss}_{xx} - \nn^\mathrm{ss}_{zz}<0$, 
albeit decreasing in amplitude when $\phi$ increases.
Interestingly, the value of $\nn^\mathrm{ss}_{xx} - \nn^\mathrm{ss}_{zz}$ becomes positive for small enough $\epsilon_\mathrm{c}$, which highlights how subtle the hydrodynamic stabilization is.

We performed shear rotations experimentally, numerically, and in a constitutive model, measuring both the shear viscosity $\eta_{12}$ and the orthogonal viscosity $\eta_{32}$ which is not measurable in a conventional rotational rheometer. 
It revealed a rich phenomenology.
The shear viscosity exhibits a dip (except at small $\theta$ for the smallest $\phi$ explored here, $\phi=0.45$) on strain scales of order unity or less, and its amplitude increases upon increase of $|\theta|$ or $\phi$. 
Remarkably, we find that $\eta_{32} \neq 0$ during the post-rotation transient, with $|\eta_{32}|$ reaching up to $\SI{50}{\percent}$ of $\eta_{12}$ at $\phi = 0.57$, and $\SI{10}{\percent}$ of $\eta_{12}^{\mathrm{SS}}$.
The qualitative angular structure of $\eta_{32}(\gamma, \theta)$ depends on $\phi$, which is explained by the predominance of either hydrodynamic or contact stresses.
For $\phi=0.45$, the second harmonic of $\eta(\gamma, \theta)$ is large as hydrodynamic stresses are significant, while for $\phi=0.57$ it is small as contact stresses dominate.
Consequently, for small $\theta$, $\eta_{32}$ produces a force that acts to reduce (stabilize) $\theta$ at smaller $\phi$ while it increases (destabilizes) $\theta$ at larger $\phi$.

In an actual non-uniform or unsteady flow where strain axes rotation occur over a strain $\gamma_\mathrm{unsteady}$, we can define a Deborah-like number $\mathrm{De} = \gamma_\mathrm{min}/\gamma_\mathrm{unsteady}$~\cite{macoskoRheologyPrinciplesMeasurements1994,poole2012deborah}, such that one should expect to observe the transient effects described here when $\mathrm{De} \gtrsim 1$.
The decrease of $\eta_{12}$ under shear rotations has already been 
used to suggest 
energy-saving flow strategies~\cite{lin_tunable_2016,ness_shaken_2018,acharyaOptimumDissipationCruising2023}, 
however the behavior of $\eta_{32}$ has so far been overlooked.
Whereas in this work we impose the deformation and measure $\eta_{32}$, 
in many cases one imposes the force or stress. 
A finite $\eta_{32}$ may lead to non-trivial trajectories, e.g. during the pulling or the sedimentation of an object in a dense suspension, especially if the suspension is not stable against shear rotations.

We have focused on a subset of the possible flow changes in a complex geometry. 
One would also need to characterize changes from simple shear to extensional flows and non-uniform flows, which are known to induce migration phenomena~\cite{guazzelliRheologyDenseGranular2018}.
While characterizing all flow histories relevant for applications (or even a carefully selected subset) is a major task, 
our results show that it would certainly reveal 
non-trivial yet possibly important stress responses.
The GW model captures the salient features of the stress response under shear rotation and in non-uniform flows~\cite{gillissenModelingMicrostructureStress2020}, and could also prove an efficient design tool in this endeavour.

\paragraph{Acknowledgements.}

Work funded in part by the European Research Council under the Horizon 2020 Programme, ERC grant agreement number 740269.


\nocite{Jeffrey_1984,Ball_1997,Cundall_1979,Brady_1988,arshad2021experimental,bounoua2019shear}


%


\end{document}